\documentclass[preprint,amsmath,amssymb,aps,showkeys,showpacs]{revtex4}
\usepackage[english]{babel}
\usepackage{graphicx}
\usepackage{graphics}
\usepackage{amsmath}
\usepackage{dcolumn}
\usepackage{amssymb}
\usepackage{bm}

\begin{document}
\title{Topological and rotating effects on the Dirac field in the spiral dislocation spacetime}
\author{A. V. D. M. Maia}
\affiliation{Departamento de F\'isica, Universidade Federal da Para\'iba, Caixa Postal 5008, 58051-900, Jo\~ao Pessoa, PB, Brazil.}

\author{K. Bakke}
\email{kbakke@fisica.ufpb.br}
\affiliation{Departamento de F\'isica, Universidade Federal da Para\'iba, Caixa Postal 5008, 58051-900, Jo\~ao Pessoa, PB, Brazil.}

\begin{abstract}

By considering a spacetime with a spiral dislocation, we analyse the behaviour of the Dirac field subject to a hard-wall confining potential. In search of relativistic bound states solutions, we discuss the influence of the topology of the spiral dislocation spacetime on the energy levels. Further, we analyse the effects of rotation on the Dirac field in the spiral dislocation spacetime. We show that both rotation and the topology of the spacetime impose a restriction on the values of the radial coordinate. Thus, we analyse the effects of rotation and the topology of the spiral dislocation spacetime on the Dirac field subject to a hard-wall confining potential by searching for relativistic bound states solutions.

\end{abstract}


\maketitle

\section{Introduction}

In recent decades, topological defect spacetimes have brought a great interest in quantum field theory and gravitation. Interesting points have been raised with respect to quantum effects in different scenarios. These quantum effects are associated with the influence of a topological defect spacetime, for instance, on the geometric quantum phase \cite{valdir4} or on the spectrum of energy of bound states \cite{valdir}. Since then, a great deal of work has dealt with topological defects in spacetime \cite{valdir2,valdir3,fur,vb2,bf,bf2,1,2,3,4,5,6,7,8,9,10,11,12,13,14,15,16,17,18,19,20,21,22,23,24,25,26,27,28,29,30,31,g5,g6,g7,g1,g2,g3,g4}. Another point of view was shown in Refs. \cite{kleinert,kat}, where topological defects in solids can be described by the Riemann-Cartan geometry. Thereby, these defects in solids become described by the spacial part of the line element of a topological defect spacetime \cite{val,put}. In the perspective of searching for analogue effects of the Aharonov-Bohm effect \cite{ab,pesk}, several works have shown the possibility of finding them in solids \cite{fur2,fur3,fur4}.

Besides, quantum systems in topological defect spacetimes have been investigated in the presence of a uniformly rotating frame \cite{vb,b,rot1,rot2,rot3,rot5}. These works have dealt with rotation in the spacetime based on the discussion raised by Landau and Lifshitz \cite{landau3}. In short, Landau and Lifshitz showed that the Minkowski spacetime has a singular behaviour at larges distances in a uniformly rotating frame. This can be viewed by performing a coordinate transformation $\varphi\rightarrow\varphi+\omega\,t$, where $\omega$ is the constant angular velocity of the rotating frame. In this way, since the line element of the Minkowski spacetime in cylindrical coordinates is given by $ds^{2}=-c^{2}\,dt^{2}+dr^{2}+r^{2}\,d\varphi^{2}+dz^{2}$, then, with $\varphi\rightarrow\varphi+\omega\,t$, we have
\begin{eqnarray}
ds^{2}=-c^{2}\left(1-\frac{\omega^{2}\,r^{2}}{c^{2}}\right)dt^{2}+2\omega\,r^{2}\,d\varphi\,dt+dr^{2}+r^{2}\,d\varphi^{2}+dz^{2}.
\label{a}
\end{eqnarray}
Hence, we can see in the line element (\ref{a}) that the radial coordinate becomes restricted to the values:
\begin{eqnarray}
0\,<\,r\,<\frac{c}{\omega}.
\label{b}
\end{eqnarray}

The meaning of this restriction is that if $r\geq\frac{c}{\omega}$ the line element (\ref{a}) becomes positive, therefore, it is not admissible. Thereby, the particle would have a velocity greater than the velocity of light if $r\geq\frac{c}{\omega}$. At present days, effects of rotation on a Dirac particle have been investigated in the Minkowski spacetime in Refs. \cite{b3,rot4,rot6}. In the cosmic string spacetime, rotating effects have been investigated in Refs. \cite{b,rot1,rot2,rot3,rot5}.

In this work, we analyse the influence of the topology of a spacetime with a spiral dislocation on the Dirac field. We deal with the Dirac field subject to a hard-wall confining potential in the spiral dislocation spacetime. Further, we analyse rotating effects on the Dirac particle in the spiral dislocation spacetime. In both cases, we focus on particular cases where a discrete spectrum of energy can be obtained.

The structure of this paper is: in section II, we introduce the line element of the spiral dislocation spacetime. Then, we confine the Dirac particle to a hard-wall confining potential and analyse the effects of the topology of the spacetime on the relativistic energy levels; in section III, we analyse the effects of rotation and topology on the Dirac particle by confining it to a hard-wall confining potential; in section IV, we present our conclusions.

\section{Spacetime with a spiral dislocation}

In this section, we analyse a Dirac particle subject to a hard-wall confining potential in spacetime with a spiral dislocation \cite{val,vb,bf}. The spacial part of this linear topological defect spacetime corresponds to the distortion of a circle into a spiral. We shall use the units $\hbar=1$ and $c=1$, therefore, the spiral dislocation spacetime is described the line element \cite{val,bf,mb}:
\begin{eqnarray}
ds^{2}=-dt^{2}+dr^{2}+2\beta\,dr\,d\varphi+\left(\beta^{2}+r^{2}\right)d\varphi^{2}+dz^{2},
\label{1.1}
\end{eqnarray}
where the constant $\beta$ is the parameter related to the distortion of the topological defect.

With the purpose of analysing the Dirac particle in the spiral dislocation spacetime, it is convenient to deal with spinors in curved spacetime background as discussed in Ref. \cite{bd}. Hence, spinors are defined in the local reference frame of the observers via a non-coordinate basis $\hat{\theta}^{a}=e^{a}_{\,\,\,\mu}\left(x\right)\,dx^{\mu}$. The indices $a,b,c=0,1,2,3$ indicate the local reference frame. The components $e^{a}_{\,\,\,\mu}\left(x\right)$ are called \textit{tetrads} and satisfy the relation \cite{,weinberg,bd,naka}:
\begin{eqnarray}
g_{\mu\nu}\left(x\right)=e^{a}_{\,\,\,\mu}\left(x\right)\,e^{b}_{\,\,\,\nu}\left(x\right)\,\eta_{ab}.
\label{1.2}
\end{eqnarray}
Note that $\eta_{ab}=\mathrm{diag}(- + + +)$ is the Minkowski tensor. Besides, the tetrads have an inverse, where we can define it by $dx^{\mu}=e^{\mu}_{\,\,\,a}\left(x\right)\,\hat{\theta}^{a}$. They are related through $e^{a}_{\,\,\,\mu}\left(x\right)\,e^{\mu}_{\,\,\,b}\left(x\right)=\delta^{a}_{\,\,\,b}$ and $e^{\mu}_{\,\,\,a}\left(x\right)\,e^{a}_{\,\,\,\nu}\left(x\right)=\delta^{\mu}_{\,\,\,\nu}$. With these definitions, we can write the tetrads and the inverse as \cite{bf}
\begin{eqnarray}
e^{a}_{\,\,\,\mu}\left(x\right)=\left(
\begin{array}{cccc}
1 & 0 & 0 & 0 \\
0 & 1 & \beta & 0 \\
0 & 0 & r & 0\\
0 & 0 & 0 & 1\\	
\end{array}\right);\,\,\,\,\,\,\,\,\,
e^{\mu}_{\,\,\,a}\left(x\right)=\left(
\begin{array}{cccc}
1 & 0 & 0 & 0 \\
0 & 1 & -\frac{\beta}{r} & 0 \\
0 & 0 & \frac{1}{r} & 0\\
0 & 0 & 0 & 1\\	
\end{array}\right)
\label{1.3}
\end{eqnarray}

This choice of the tetrads allows us to solve the Maurer-Cartan structure equations \cite{naka}: $T^{a}=d\hat{\theta}^{a}+\omega^{a}_{\,\,\,b}\wedge\hat{\theta}^{b}$. The operator $d$ corresponds to the exterior derivative and the symbol $\wedge$ means the wedge product. Moreover, $T^{a}=T^{a}_{\,\,\,\mu\nu}\,dx^{\mu}\wedge dx^{\nu}$ corresponds to the torsion 2-form, while $\omega^{a}_{\,\,\,b}=\omega_{\mu\,\,\,\,b}^{\,\,\,a}\left(x\right)\,dx^{\mu}$ corresponds to the connection 1-form. Then, by solving the Maurer-Cartan structure equations, we have \cite{bf}
\begin{eqnarray}
T^{1}=2\pi\,\beta\,\delta\left(r\right)\,\delta\left(\varphi\right)\,dr\wedge d\varphi;\,\,\,\,\,\,\,\,\omega_{\varphi\,\,\,\,1}^{\,\,\,\,2}\left(x\right)=-\omega_{\varphi\,\,\,\,\,2}^{\,\,\,\,1}\left(x\right)=1.
\label{1.4}
\end{eqnarray}

In curved spacetime background and in the presence of torsion, the Dirac equation is written in terms of the covariant derivative, which in turn is defined as 
\cite{t10,bf}:
\begin{eqnarray}
\widetilde{\nabla}_{\mu}=\partial_{\mu}+\frac{i}{4}\,\omega_{\mu ab}\left(x\right)\Sigma^{ab}+\frac{i}{4}K_{\mu ab}\left(x\right)\Sigma^{ab}.
\label{1.5}
\end{eqnarray}
The term $\Gamma_{\mu}=\frac{i}{4}\,\omega_{\mu ab}\left(x\right)\,\Sigma^{ab}$ is the spinorial connection \cite{bd,naka} and $\Sigma^{ab}=\frac{i}{2}\left[\gamma^{a},\gamma^{b}\right]$. The $\gamma^{a}$ matrices are defined in the local reference frame. They correspond to the standard Dirac matrices in the Minkowski spacetime \cite{greiner}:
\begin{eqnarray}
\gamma^{0}=\hat{\beta}=\left(
\begin{array}{cc}
1 & 0 \\
0 & -1 \\
\end{array}\right);\,\,\,\,\,\,
\gamma^{i}=\hat{\beta}\,\hat{\alpha}^{i}=\left(
\begin{array}{cc}
 0 & \sigma^{i} \\
-\sigma^{i} & 0 \\
\end{array}\right);\,\,\,\,\,\,\Sigma^{i}=\left(
\begin{array}{cc}
\sigma^{i} & 0 \\
0 & \sigma^{i} \\	
\end{array}\right).
\label{1.6}
\end{eqnarray}
Note that $\vec{\Sigma}$ is the spin vector. Furthermore, the matrices $\sigma^{i}$ correspond to the Pauli matrices, and thus, satisfy the relation $\left(\sigma^{i}\,\sigma^{j}+\sigma^{j}\,\sigma^{i}\right)=2\,\eta^{ij}$.  The indices $(i,j,k=1,2,3)$ indicate the spatial indices of the local reference frame.  We should observe that the last term of Eq. (\ref{1.5}) is defined in terms of the object $K_{\mu ab}\left(x\right)$, which is related to the contortion tensor by \cite{t10}
\begin{eqnarray}
K_{\mu ab}\left(x\right)= K_{\beta\nu\mu}\left(x\right)\left[e^{\nu}_{\,\,\,a}\left(x\right)\,e^{\beta}_{\,\,\,b}\left(x\right)-e^{\nu}_{\,\,\,b}\left(x\right)\,e^{\beta}_{\,\,\,a}\left(x\right)\right].
\label{1.7}
\end{eqnarray}
The contortion tensor, in turn, is related to the torsion tensor via
\begin{eqnarray}
K^{\beta}_{\,\,\,\nu\mu}=\frac{1}{2}\left(T^{\beta}_{\,\,\,\nu\mu}-T_{\nu\,\,\,\,\mu}^{\,\,\,\beta}-T^{\,\,\,\beta}_{\mu\,\,\,\,\nu}\right).
\label{1.8}
\end{eqnarray}

By following Ref. \cite{t10}, the torsion tensor is antisymmetric in the last two indices, while the contortion tensor is antisymmetric in the first two indices. In this way, we can rewrite the torsion tensor in terms of three irreducible components:
\begin{eqnarray}
T_{\mu}=T^{\beta}_{\,\,\,\mu\beta};\,\,\,\,\,\,S^{\alpha}=\epsilon^{\alpha\beta\nu\mu}\,T_{\beta\nu\mu},
\label{1.9}
\end{eqnarray}
and in the tensor $q_{\beta\nu\mu}$ that satisfies the conditions: $q^{\beta}_{\,\,\mu\beta}=0$ and $\epsilon^{\alpha\beta\nu\mu}\,q_{\beta\nu\mu}=0$. Thus, the torsion tensor becomes: $T_{\alpha\beta\mu}=\frac{1}{3}\left(T_{\beta}\,g_{\alpha\mu}\left(x\right)-T_{\mu}\,g_{\alpha\beta}\left(x\right)\right)-\frac{1}{6}\epsilon_{\alpha\beta\mu\nu}\,S^{\nu}+q_{\alpha\beta\mu}$.

As shown in Ref. \cite{bf}, by taking the tetrads (\ref{1.3}) and the torsion $2$-form (\ref{1.4}), we obtain that the only non-null component of the irreducible components of the torsion tensor is
\begin{eqnarray}
T_{\varphi}=T_{\,\,\,\varphi\,r}^{r}=-T_{\,\,\,r\,\varphi}^{r}=-2\pi\,\chi\,\delta\left(r\right)\delta\left(\varphi\right).
\label{1.10}
\end{eqnarray}
This is a component of the trace four-vector $T_{\mu}$. An interesting point raised by Ref. \cite{t10} is that the trace four-vector $T_{\mu}$ and the tensor $q_{\beta\nu\mu}$ decouples with fermions. On the other hand, the axial four-vector $S^{\alpha}$ couples with fermions. Therefore, as discussed in Refs. \cite{t10,bf}, the trace four-vector $T_{\mu}$ is introduced into the Dirac equation through a nonminimal coupling given by:
\begin{eqnarray}
i\gamma^{\mu}\,\nabla_{\mu}\rightarrow i\gamma^{\mu}\,\nabla_{\mu}+\nu\,\gamma^{\mu}\,T_{\mu},
\label{1.11}
\end{eqnarray}
where $\nu$ is an arbitrary nonminimal coupling parameter (dimensionless) and $\nabla_{\mu}=\partial_{\mu}+\frac{i}{4}\,\omega_{\mu ab}\left(x\right)\Sigma^{ab}$. By using the tetrads field (\ref{1.3}), the Dirac equation becomes
\begin{eqnarray}
m\psi=i\gamma^{0}\frac{\partial\psi}{\partial t}+i\gamma^{1}\left(\frac{\partial}{\partial\,r}+\frac{1}{2r}\right)\psi+i\frac{\gamma^{2}}{r}\left(\frac{\partial}{\partial\varphi}-\beta\frac{\partial}{\partial\,r}-i\nu\,T_{\varphi}\right)\psi+i\gamma^{3}\frac{\partial\psi}{\partial z}.
\label{1.12}
\end{eqnarray}

Note that, for $r\neq0$, the contribution to the Dirac equation associated with $T_{\mu}$ vanishes. In this way, for $r\neq0$, the Dirac equation (\ref{1.12}) becomes
\begin{eqnarray}
i\frac{\partial\psi}{\partial t}=m\hat{\beta}\psi-i\hat{\alpha}^{1}\left(\frac{\partial}{\partial\,r}+\frac{1}{2r}\right)\psi-i\frac{\hat{\alpha}^{2}}{r}\left(\frac{\partial}{\partial\varphi}-\beta\frac{\partial}{\partial\,r}\right)\psi-i\hat{\alpha}^{3}\frac{\partial\psi}{\partial z}.
\label{1.13}
\end{eqnarray} 
The solution to the Dirac equation (\ref{1.13}) can be given in the form:
\begin{eqnarray}
\psi=e^{-i\mathcal{E}t}\,\left(
\begin{array}{c}
\phi\\
\xi\\	
\end{array}\right),
\label{1.14}
\end{eqnarray} 
where $\phi=\phi\left(r,\varphi,z\right)$ and $\xi=\xi\left(r,\varphi,z\right)$ are two-spinors. By substituting (\ref{1.14}) into the Dirac equation (\ref{1.13}), we obtain two coupled equations of $\phi$ and $\chi$. The first coupled equation is
\begin{eqnarray}
\left(\mathcal{E}-m\right)\phi=\left[-i\sigma^{1}\frac{\partial}{\partial r}-\frac{i\sigma^{1}}{2r}+i\frac{\beta\,\sigma^{2}}{r}\frac{\partial}{\partial r}-\frac{i\sigma^{2}}{r}\frac{\partial}{\partial\varphi}-i\sigma^{3}\frac{\partial}{\partial z}\right]\xi,
\label{1.15}
\end{eqnarray}
while the second coupled equation is 
\begin{eqnarray}
\left(\mathcal{E}+m\right)\xi=\left[-i\sigma^{1}\frac{\partial}{\partial r}-\frac{i\sigma^{1}}{2r}+i\frac{\beta\,\sigma^{2}}{r}\frac{\partial}{\partial r}-\frac{i\sigma^{2}}{r}\frac{\partial}{\partial\varphi}-i\sigma^{3}\frac{\partial}{\partial z}\right]\phi.
\label{1.16}
\end{eqnarray}
Then, by eliminating $\xi$ in Eq. (\ref{1.16}) and by substituting it into Eq. (\ref{1.15}), we obtain the following second-order differential equation:
\begin{eqnarray}
\left(\mathcal{E}^{2}-m^{2}\right)\phi&=&-\left[1+\frac{\beta^{2}}{r^{2}}\right]\frac{\partial^{2}\phi}{\partial r^{2}}-\left[\frac{1}{r}-\frac{\beta^{2}}{r^{3}}\right]\frac{\partial\phi}{\partial r}-\frac{1}{r^{2}}\frac{\partial^{2}\phi}{\partial\varphi^{2}}+\frac{i\sigma^{3}}{r^{2}}\frac{\partial\phi}{\partial\varphi}+\frac{1}{4r^{2}}\,\phi\nonumber\\
[-2mm]\label{1.17}\\[-2mm]
&-&\frac{i\beta\,\sigma^{3}}{r^{2}}\frac{\partial\phi}{\partial r}+\frac{i\beta\,\sigma^{3}}{2r^{3}}\,\phi-\frac{\beta}{r^{3}}\frac{\partial\phi}{\partial\varphi}+\frac{2\beta}{r^{2}}\frac{\partial^{2}\phi}{\partial\varphi\partial r}-\frac{\partial^{2}\phi}{\partial z^{2}}.\nonumber
\end{eqnarray}

Observe that the solution to Eq. (\ref{1.17}) can be written in terms of the eigenvalues of the $z$-component of the total angular momentum and the linear momentum operators: $\phi\left(r,\,\varphi,\,z\right)=e^{i\left(l+1/2\right)\varphi+ikz}\,u\left(r\right)$, where $k$ is a constant and $l=0,\pm1,\pm2,\pm3,\pm4\ldots$. Besides, we have that $\sigma^{3}\phi=\pm\phi=s\phi$. Henceforth, let us take $k=0$. Then, by substituting this solution into Eq. (\ref{1.17}), we obtain the following radial equation: 
\begin{eqnarray}
\left(1+\frac{\beta^{2}}{r^{2}}\right)u''+\left(\frac{1}{r}-\frac{\beta^{2}}{r^{3}}-i\frac{2\beta\,\zeta}{r^{2}}\right)u'-\frac{\zeta^{2}}{r^{2}}\,u+i\frac{\beta\,\zeta}{r^{3}}\,u+\tau^{2}\,u=0,
\label{1.18}
\end{eqnarray}
where we have defined the parameters:
\begin{eqnarray}
\zeta=l+\frac{1}{2}\left(1-s\right);\,\,\,\,\tau^{2}=\mathcal{E}^{2}-m^{2}.
\label{1.19}
\end{eqnarray}

In search of a solution to the radial equation (\ref{1.18}), let us write \cite{mb,vb}:
\begin{eqnarray}
u\left(r\right)=\exp\left(i\,\zeta\,\tan^{-1}\left(\frac{r}{\beta}\right)\right)\times f\left(r\right),
\label{1.20}
\end{eqnarray}
where $f\left(r\right)$ is an unknown function. Then, by substituting the radial wave function (\ref{1.20}) into Eq. (\ref{1.18}), we obtain 
\begin{eqnarray}
\left(1+\frac{\beta^{2}}{r^{2}}\right)f''+\left(\frac{1}{r}-\frac{\beta^{2}}{r^{3}}\right)f'-\frac{\zeta^{2}}{\left(r^{2}+\beta^{2}\right)}\,f+\tau^{2}\,f=0.
\label{1.21}
\end{eqnarray}
Further, let us define $x=\tau\sqrt{r^{2}+\beta^{2}}$, and thus, we can rewrite Eq. (\ref{1.21}) in the form:
\begin{eqnarray}
f''+\frac{1}{x}\,f'-\frac{\zeta^{2}}{x^{2}}\,f+\frac{1}{4}\,f=0.
\label{1.22}
\end{eqnarray}

Hence, Eq. (\ref{1.22}) corresponds to the Bessel equation \cite{arf,abra}. Its general solution is given by
\begin{eqnarray}
f\left(x\right)=A\,J_{\left|\zeta\right|}\left(x\right)+B\,N_{\left|\zeta\right|}\left(x\right),
\label{1.23}
\end{eqnarray} 
where $J_{\left|\zeta\right|}\left(x\right)$ and $N_{\left|\zeta\right|}\left(x\right)$ are the Bessel functions of first and second kinds \cite{abra,arf}. A regular solution at the origin is obtained by taking $B=0$ in Eq. (\ref{1.23}), because the function $N_{\left|\zeta\right|}\left(x\right)$ diverges at the origin. Then, the solution to Eq. (\ref{1.22}) becomes
\begin{eqnarray}
f\left(x\right)=A\,J_{\left|\zeta\right|}\left(x\right),
\label{1.24}
\end{eqnarray}
where $A$ is a constant. In the following, let us consider the Dirac particle to be confined to a hard-wall confining potential. This confinement imposes that the wave function vanishes at a fixed value of the radial coordinate $r_{0}$. Therefore, the function $f\left(x\right)$ must vanish when $x\rightarrow x_{0}=\tau\sqrt{r^{2}_{0}+\beta^{2}}$: 
\begin{eqnarray}
f\left(x_{0}\right)=0.
\label{1.25}
\end{eqnarray}
Let us also assume that $x_{0}\gg1$. In this way, we have \cite{arf,abra}:
\begin{eqnarray}
J_{\left|\zeta\right|}\left(x_{0}\right)\rightarrow\sqrt{\frac{2}{\pi\,x_{0}}}\,\cos\left(x_{0}-\frac{\left|\zeta\right|\,\pi}{2}-\frac{\pi}{4}\right).
\label{1.26}
\end{eqnarray}

Therefore, by substituting Eqs. (\ref{1.26}) and (\ref{1.24}) into Eq. (\ref{1.25}), we obtain
\begin{eqnarray}
\mathcal{E}_{n,\,l,\,s}\approx\pm\sqrt{m^{2}+\frac{\pi^{2}}{\left(r_{0}^{2}+\beta^{2}\right)}\left[n+\frac{\left|\zeta\right|}{2}+\frac{3}{4}\right]^{2}},
\label{1.27}
\end{eqnarray}
where $n=0,1,2,\ldots$ is the quantum number related to the radial modes.

Hence, the relativistic energy levels given in Eq. (\ref{1.27}) are achieved when the Dirac particle is confined to a hard-wall confining potential in the spiral dislocation spacetime. The effects that stem from the topology of the spacetime is given by the presence of the effective radius $\rho_{0}=\sqrt{r_{0}^{2}+\beta^{2}}$. Besides, there is no analogue of the Aharonov-Bohm effect for bound states \cite{pesk}. As shown in Ref. \cite{fur}, this effect is determined by a shift in the angular momentum quantum number yielded by the topology of the spacetime. In the present case, the angular momentum is given by $\zeta=l+\frac{1}{2}\left(1-s\right)$, i.e., it remains unchanged under the influence of the topology of the spiral dislocation spacetime. Moreover, if we take $\beta=0$, we obtain the relativistic energy levels for a Dirac particle confined to a hard-wall confining potential in the Minkowski spacetime.

Finally, let us discuss the non-relativistic limit of the spectrum of energy (\ref{1.27}). By applying the binomial expansion up to terms of order $m^{-1}$, we thus obtain
\begin{eqnarray}
\mathcal{E}_{n,\,l,\,s}&\approx&\frac{\pi^{2}}{2m\left(r_{0}^{2}+\beta^{2}\right)}\left[n+\frac{\left|\zeta\right|}{2}+\frac{3}{4}\right]^{2}.
\label{1.28}
\end{eqnarray}

Therefore, we have in Eq. (\ref{1.28}) the spectrum of energy of a non-relativistic Dirac particle confined to a hard-wall confining potential in the presence of a spiral dislocation. Even in the non-relativistic limit, the contribution to the energy levels that stems from the topology of the defects is given by the effective radius $\rho_{0}=\sqrt{r_{0}^{2}+\beta^{2}}$. In the same way of the relativistic case, there is no Aharonov-Bohm-type effect for bound states in the non-relativistic limit.


\section{Rotating reference frame}

In this section, we shall consider a uniformly rotating frame, and thus, analyse the effects associated with rotation and the topology of the spiral dislocation spacetime. Let us perform a coordinate transformation given by: $\varphi\rightarrow\varphi+\omega\,t$. Thereby, the line element (\ref{1.1}) becomes \cite{vb}:
\begin{eqnarray}
ds^{2}&=&-\left(1-\omega^{2}\,\beta^{2}-\omega^{2}\,r^{2}\right)dt^{2}+2\beta\,\omega\,dr\,dt+2\omega\left(\beta^{2}+r^{2}\right)d\varphi\,dt\nonumber\\
[-2mm]\label{2.1}\\[-2mm]
&+&dr^{2}+2\beta\,dr\,d\varphi+\left(\beta^{2}+r^{2}\right)d\varphi^{2}+dz^{2}.\nonumber
\end{eqnarray}

As shown in Ref. \cite{landau3} in the Minkowski spacetime, in a uniformly rotating frame, the line element is not well-defined for large distances. In the present case, the line element (\ref{2.1}) shows us the same behaviour. From Eq. (\ref{2.1}), the radial coordinate is restricted by range:
\begin{eqnarray}
0\,\leq\,r\,<\frac{\sqrt{1-\beta^{2}\omega^{2}}}{\omega}.
\label{2.2}
\end{eqnarray}
Hence, the restriction on the radial coordinate is determined by the angular velocity and the parameter associated with torsion in the spacetime. Therefore, if $r\,\geq\,\frac{\sqrt{1-\beta^{2}\omega^{2}}}{\omega}$, we would have the particle to be placed outside of the light-cone. Note that we recover the Minkowski spacetime case discussed in Ref. \cite{landau3} by taking $\beta=0$.

Since, we want to analyse topological and rotating effects on the Dirac particle in the spiral dislocation spacetime, let us define the tetrads and its inverse as follows:
\begin{eqnarray}
e^{a}_{\,\,\,\mu}\left(x\right)=\left(
\begin{array}{cccc}
1 & 0 & 0 & 0 \\
\omega\beta & 1 & \beta & 0 \\
\omega\,r & 0 & r & 0\\
0 & 0 & 0 & 1\\	
\end{array}\right);\,\,\,\,\,\,\,\,\,
e^{\mu}_{\,\,\,a}\left(x\right)=\left(
\begin{array}{cccc}
1 & 0 & 0 & 0 \\
0 & 1 & -\frac{\beta}{r} & 0 \\
-\omega & 0 & \frac{1}{r} & 0\\
0 & 0 & 0 & 1\\	
\end{array}\right).
\label{2.3}
\end{eqnarray}
Thus, by solving the Maurer-Cartan structure equations \cite{naka}, we have
\begin{eqnarray}
T^{1}&=&2\pi\,\beta\,\delta\left(\rho\right)\,\delta\left(\varphi\right)\,d\rho\wedge d\varphi;\nonumber\\
\omega_{\varphi\,\,\,\,1}^{\,\,\,\,2}\left(x\right)&=&-\omega_{\varphi\,\,\,\,\,2}^{\,\,\,\,1}\left(x\right)=1;\label{2.4}\\
\omega_{t\,\,\,\,1}^{\,\,\,\,2}\left(x\right)&=&-\omega_{t\,\,\,\,\,2}^{\,\,\,\,1}\left(x\right)=\omega.\nonumber
\end{eqnarray}

Hence, by using (\ref{2.3}) and (\ref{2.4}), the Dirac equation becomes 
\begin{eqnarray}
i\frac{\partial\psi}{\partial t}=m\hat{\beta}\psi+i\omega\,\frac{\partial\psi}{\partial\varphi}-i\hat{\alpha}^{1}\left(\frac{\partial}{\partial\,r}+\frac{1}{2r}\right)\psi-i\frac{\hat{\alpha}^{2}}{r}\left(\frac{\partial}{\partial\varphi}-\beta\frac{\partial}{\partial\,r}\right)\psi-i\hat{\alpha}^{3}\frac{\partial\psi}{\partial z},
\label{2.4a}
\end{eqnarray} 
and thus, by following the steps from Eq. (\ref{1.14}) to Eq. (\ref{1.19}), we obtain
\begin{eqnarray}
\left(1+\frac{\beta^{2}}{r^{2}}\right)u''+\left(\frac{1}{r}-\frac{\beta^{2}}{r^{3}}-i\frac{2\beta\,\zeta}{r^{2}}\right)u'-\frac{\zeta^{2}}{r^{2}}\,u+i\frac{\beta\,\zeta}{r^{3}}\,u+\theta^{2}\,u=0,
\label{2.5}
\end{eqnarray}
where 
\begin{eqnarray}
\theta^{2}=\left[\mathcal{E}+\omega\left(l+1/2\right)\right]^{2}-m^{2}.
\label{2.6}
\end{eqnarray}

Then, we can also follow the steps from Eq. (\ref{1.20}) to Eq. (\ref{1.24}) by defining $y=\theta\sqrt{\left(r^{2}+\beta^{2}\right)}$. Since the radial coordinate is restricted to the values defined in Eq. (\ref{2.2}), hence, the radial wave function must be normalized inside the physical region of the spacetime determined in Eq. (\ref{2.2}). Thereby, let us impose that the radial wave function vanishes when $r\rightarrow\frac{\sqrt{1-\omega^{2}\beta^{2}}}{\omega}$. From this perspective, we can define $r_{0}=\frac{\sqrt{1-\omega^{2}\beta^{2}}}{\omega}$, and thus, $y_{0}=\theta\sqrt{r^{2}_{0}+\beta^{2}}=\theta/\omega$. Therefore, the boundary condition (\ref{1.25}) becomes
\begin{eqnarray}
f\left(y\rightarrow y_{0}=\theta/\omega\right)=0.
\label{2.7}
\end{eqnarray}
By considering $y_{0}\gg1$, we can also use the relation (\ref{1.26}). Thus, by following the steps from (\ref{1.24}) to Eq. (\ref{1.27}), we obtain
\begin{eqnarray}
\mathcal{E}_{n,\,l,\,s}\approx-\omega\left(l+1/2\right)\pm\sqrt{m^{2}+\pi^{2}\omega^{2}\left[n+\frac{\left|\zeta\right|}{2}+\frac{3}{4}\right]^{2}},
\label{2.8}
\end{eqnarray}
where $n=0,1,2,\ldots$ is also the quantum number related to the radial modes.

Hence, the spectrum of energy (\ref{2.8}) is the relativistic energy levels for a Dirac particle confined to a hard-wall confining potential in the spiral dislocation spacetime under the effects of rotation. In this case, since the physical region of the spacetime is defined in the range $0\,\leq\,r\,<\frac{\sqrt{1-\beta^{2}\omega^{2}}}{\omega}$, therefore, the geometry of the spacetime has played the role of the hard-wall confining potential. Observe that there is no influence of the topology of the spiral dislocation spacetime on the relativistic energy levels (\ref{2.8}). Despite having effects of the topology of the spacetime on the radial coordinate as shown in Eq. (\ref{2.2}), there are only the effects of rotation on the relativistic spectrum of energy (\ref{2.8}). It is analogous to the case of the Minkowski spacetime analysed in Ref. \cite{b3}. This behaviour has also been seen for the scalar field in the spiral dislocation spacetime under the effects of rotation \cite{vb}. Furthermore, due to the effects of rotation, there is a contribution to the relativistic energy levels that corresponds to the coupling between the angular velocity $\omega$ and the angular momentum $l$ given in the first term of Eq. (\ref{2.8}). It corresponds to a Sagnac-type effect \cite{r4}. By comparing with the relativistic energy levels (\ref{1.27}) obtained in the absence of rotation, we can observe that the (total) angular momentum $\zeta=l+\frac{1}{2}\left(1-s\right)$ remains unchanged under the influence of rotation and topology of the spiral dislocation spacetime. Therefore, there is no Aharonov-Bohm-type effect for bound states.

Finally, let us discuss the non-relativistic limit of the energy levels (\ref{2.8}). By applying the binomial expansion up to terms of order $m^{-1}$ in Eq. (\ref{2.8}), we have
\begin{eqnarray}
\mathcal{E}_{n,\,l,\,s}\approx\frac{\pi^{2}\omega^{2}}{2m}\left[n+\frac{\left|\zeta\right|}{2}+\frac{3}{4}\right]^{2}-\omega\left(l+1/2\right).
\label{2.9}
\end{eqnarray}

Then, the spectrum of energy (\ref{2.9}) stems from the confinement of a non-relativistic Dirac particle to a hard-wall confining potential in the spiral dislocation spacetime under the effects of rotation. We can also observe in the non-relativistic case that there is no influence of the topology of the spiral dislocation spacetime on the spectrum of energy. It is analogous to the spectrum of energy in the absence of defect obtained in Ref. \cite{b3}. The effects of rotation also yield the coupling between the angular velocity $\omega$ and the angular momentum $l$ given in the last term of Eq. (\ref{2.9}). This coupling is known as the Page-Werner {\it et al} term \cite{r1,r2,r4}. Besides, in contrast to the non-relativistic energy levels (\ref{1.28}) obtained in the absence of rotation, we have that the (total) angular momentum $\zeta=l+\frac{1}{2}\left(1-s\right)$ is not modified by the rotation and the topology of the spiral dislocation spacetime. Therefore, there is no Aharonov-Bohm-type effect for bound states.

\section{Conclusions}

We have started this work by introducing the line element of the spacetime with a spiral dislocation. Then, we have investigated the topological effects on a Dirac particle confined to a hard-wall confining potential in this spacetime background. We have obtained a discrete spectrum of energy, where it is determined by an effective radius $\rho_{0}=\sqrt{r_{0}^{2}+\beta^{2}}$ that stems from the effects of the topology of the spacetime. However, no contribution to the angular momentum quantum number is yielded by the topology of the spacetime. In this sense, there is no Aharonov-Bohm effect for bound states. Furthermore, we have analysed the non-relativistic limit of the energy levels. We have also seen that the only contribution that arises from the topological defect spacetime is given by the effective radius $\rho_{0}=\sqrt{r_{0}^{2}+\beta^{2}}$. Then, since there is no contribution to the angular momentum quantum, no Aharonov-Bohm-type effect for bound states exists.

We have gone further by analysing a Dirac particle confined to a hard-wall confining potential in the spiral dislocation spacetime under the effects of rotation. We have seen that the radial coordinate has a restriction on its values, which are determined in the range $0\,\leq\,r\,<\frac{\sqrt{1-\beta^{2}\omega^{2}}}{\omega}$. For this reason, we have imposed that the radial wave function vanishes when $r\rightarrow\frac{\sqrt{1-\omega^{2}\beta^{2}}}{\omega}$. In this sense, the geometry of the spacetime has played the role of the hard-wall confining potential. Thereby, we have obtained a discrete spectrum of energy. However,  despite having effects of the topology of the spacetime on the radial coordinate, we have seem that there are only the effects of rotation on the relativistic spectrum of energy. Therefore, the relativistic spectrum of energy is analogous to the case of the Minkowski spacetime \cite{b3}. Besides, we have observed that the coupling between the angular velocity $\omega$ and the angular momentum $l$. Hence, we have a Sagnac-type effect \cite{r4}. In the search for Aharonov-Bohm-type effects for bound states, we have seen that the (total) angular momentum $\zeta=l+\frac{1}{2}\left(1-s\right)$ is not modified by the effects of rotation and topology of the spiral dislocation spacetime. Therefore, there is no Aharonov-Bohm-type effect for bound states.

Finally, we have analysed the non-relativistic limit of the energy levels for a Dirac particle confined to a hard-wall confining potential in the spiral dislocation spacetime under the effects of rotation. We have seen no influence of the topology of the spiral dislocation spacetime on the spectrum of energy. It is analogous to the spectrum of energy in the absence of defect obtained in Ref. \cite{b3}. On the other hand, the effects of rotation also yield the coupling between the angular velocity $\omega$ and the angular momentum $l$, which is known as the Page–Werner {\it et al} term \cite{r1,r2,r4}. In search of Aharonov-Bohm-type effects for bound states, we have observed no modification in the (total) angular momentum $\zeta=l+\frac{1}{2}\left(1-s\right)$ . Therefore, in this sense, there is no Aharonov-Bohm-type effect for bound states.

\acknowledgments{The authors would like to thank CNPq for financial support.}

\end{document}